\documentclass[aps,pre,twocolumn,groupedaddress,showpacs,floatfix,suprscriptaddress]{revtex4-1}
\usepackage[utf8]{inputenc}
\usepackage[T1]{fontenc}
\usepackage{graphicx}
\usepackage{amssymb}
\usepackage{nicefrac}
\usepackage{xcolor}
\usepackage{here}

\usepackage{times}

\usepackage[colorinlistoftodos,prependcaption,textsize=small]{todonotes}

\newcommand{\bdt}[1]{\textcolor{black}{#1}}
\newcommand{\kct}[1]{\textcolor{black}{#1}}

\newcommand{\amin}{\alpha_{\mathrm{min}}}
\newcommand{\amax}{\alpha_{\mathrm{max}}}

\usepackage[inner]{showlabels}

\usepackage{amsmath}
\DeclareMathOperator{\sign}{sign}

\usepackage{hyperref}

\hypersetup{
  colorlinks=true,
citecolor=black,
linkcolor=black,
urlcolor=black
  }

\begin{document}

\title{Deterministic force-free resonant activation}

\author{Karol Capa{\l}a}
\email{karol@th.if.uj.edu.pl} \affiliation{Institute of Theoretical Physics, and Mark Kac Center for Complex Systems
Research, Jagiellonian University, ul. St. {\L}ojasiewicza 11,
30--348 Krak\'ow, Poland}

\author{Bart{\l}omiej Dybiec}
\email{bartek@th.if.uj.edu.pl} \affiliation{Institute of Theoretical Physics, and Mark Kac Center for Complex Systems
Research, Jagiellonian University, ul. St. {\L}ojasiewicza 11,
30--348 Krak\'ow, Poland}

\date{\today}

\begin{abstract}
Combined action of noise and deterministic force in dynamical systems can induce resonant effects.
Here, we demonstrate a minimal, deterministic-force-free, setup allowing for occurrence of resonant, noise induced effects.
We show that in the archetypal problem of escape from finite intervals driven by $\alpha$-stale noise with the periodically modulated stability index, depending on the initial direction of the modulation, resonant-activation-like or noise-enhanced-stability-like phenomena can be observed.
\end{abstract}

\pacs{
 05.40.Fb, % Random walks and Levy flights
 05.10.Gg, % Stochastic analysis methods (Fokker--Planck, Langevin, etc.)
 02.50.-r, % Probability theory, stochastic processes, and statistics
 02.50.Ey, % Stochastic processes
 }

\maketitle

%%%%%%%%%%%%%%%%%%%%%%%%%%%%%%%%%%%%%%%%%%%%%%%%%%%%%%%%%%%%%%%%%%%%%%%%%%%%%%%%%%%%%%%%%%%%%%%%
\section{Introduction \label{sec:intro}}

The action of a noise in the dynamical systems results in occurrence of so-called noise-induced effects \cite{horsthemke1984}, which demonstrate constructive role of fluctuations.
Combined action of stochastic and deterministic forces is responsible for emergence of many counterintuitive effects.
Among them, resonant activation (RA) \cite{devoret1984,doering1992}, stochastic resonance (SR) \cite{benzi1981,mantegna1995,gammaitoni1998}, ratcheting effect \cite{magnasco1993,reimann2002} and noise enhanced stability in \cite{agudov2001,dubkov2004,valenti2015}  are the most celebrated.
In the RA phenomenon, the escape over a modulated potential barrier under action of noise can be optimized, i.e., there exists such a parameter of barrier modulating process for which the mean first passage time is minimal.
\bdt{Efficiency of dynamic resonant activation \cite{devoret1984}, stochastic resonant activation \cite{doering1992} as well as stochastic resonance \cite{gammaitoni1995} relies on frequency or time scaling matching \cite{valenti2014,spagnolo2015,spagnolo2017nonlinear,dubkov2009}.}
Typically in the resonant activation the potential barrier is dichotomously \cite{doering1992} or periodically \cite{pankratov2000} modulated.
For the Gaussian white noise the height of the potential barrier is measured in the units of $k_BT$, therefore instead of modulating the potential barrier one can modulate the system temperature.
Here, we extend this approach by studying the escape from finite intervals under action of $\alpha$-stable noises.
In contrast to earlier studies, we assume that the stability index $\alpha$ is no longer constant, but it is altered in time.
Such a system is out of equilibrium, nevertheless changes in $\alpha$ modify the width of the noise induced displacement distribution, as measured by the interquantile distance.
In analogy to properties of L\'evy ratchets \cite{dybiec2008d,lisowski2015}, we expect that action of modulated noise can produce resonant effects in simpler setups that are traditionally considered for inspection of noise induced effects.
This hypothesis is based on non-equilibrium properties of $\alpha$-stable L\'evy type noises.

L\'evy noises are especially well suited for description of out-of-equilibrium systems, as they allow for occurrence of large fluctuations with a significantly larger probability than the Gaussian  white noise.
Well developed theory and desired mathematical properties, e.g., self similarity, infinite divisibility and generalized central limit theorem, make $\alpha$-stable noises widely applied in various out-of-equilibrium models and setups displaying anomalous fluctuations.
Non-Gaussian, heavy-tailed fluctuations have been recorded in diverse experimental setups ranging from
rotating flows \cite{solomon1993}, optical systems and materials \cite{barthelemy2008,mercadier2009levyflights}, physiological applications \cite{cabrera2004}, disordered media \cite{bouchaud1990}, biological systems \cite{bouchaud1991},
financial time series \cite{laherrere1998,mantegna2000,lera2018gross},  dispersal patterns of humans and animals \cite{brockmann2006,sims2008}, laser cooling \cite{barkai2014} to
 gaze dynamics \cite{amor2016} and search strategies \cite{shlesinger1986,reynolds2009}.
L\'evy noise system are also extensively studied theoretically \cite{getoor1961,blumenthal1961,metzler2000,barkai2001,chechkin2006,jespersen1999,klages2008,dubkov2008}.
\bdt{Furthermore, possible applications of L\'evy drivings in various systems, e.g., population dynamics \cite{Dubkov2008b} and fluctuation detectors \cite{Guarcello2019}, have been suggested.}

Models, assuming variability of system parameters, have been explored in various contexts.
For instance, the Gaussian white noise with fluctuating temperature \cite{brockmann2002} can induce L\'evy flights, which are described by the space fractional Smoluchowski-Fokker-Planck equation \cite{jespersen1999}.
Appropriate fluctuation protocol in temperature \cite{wilk2000} can transform Boltzmann-Gibbs distribution into the one following from Tsallis statistics \cite{tsallis1995,beck2001}.
In general, macroscopic fluctuations of the systems parameters are studied within superstatistics \cite{beck2003}.
Here, we assume that some of the system parameters evolve in time, but these changes are deterministic, like in a generalization of the escape from the positive half-line \cite{sparre1953,sparre1954} to the time dependent drift and diffusion coefficients  \cite{molini2011first}.
Nevertheless, the studied system displays increased randomness, because its stochastic properties are determined by the evolving parameter \cite{dybiec2008e,dybiec2009e}.

The noise driven escape from finite intervals is the archetypal problem studied within the theory of stochastic processes \cite{cox1965}.
For the Gaussian white noise it is possible to find not only the mean first passage time \cite{gardiner2009}, but also time dependent densities \cite{cox1965}.
Under the action of $\alpha$-stable noises, the mean first passage time is known \cite{getoor1961,blumenthal1961}, but time dependent densities can be constructed numerically only \cite{dybiec2010c,dybiec2012fractional}, due to difficulties in construction of eigenvalues and eigenfunctions of fractional laplacians \cite{katzav2008spectrumfractional,zozor2011spectral,kwasnicki2012eigenvalues,kirichenko2016}.
From the microscopic point of view, these problems are produced by discontinuity of trajectories of $\alpha$-stable motions.
Consequently, in order to leave the domain of motion a particle can escape via a single long jump \cite{ditlevsen1999,bier2018,imkeller2006,imkeller2006b} instead of a sequence of short jumps, which is the typical escape scenario under the Gaussian white noise driving.
Discontinuity of trajectories is also responsible for failure of method of images \cite{chechkin2003b} and leapovers of L\'evy flights \cite{koren2007,koren2007b,palyulin2019first}.

In this manuscript, we extend the discussion on the overdamped, deterministic force-free kinetics, driven by L\'evy noises.
In the next section (Sec.~\ref{sec:model} Model) we define the minimal setup allowing for occurrence of the resonant activation.
In the Sec.~\ref{sec:results} (Results) we present results of extensive  numerical simulations.
The manuscript is closed with Summary and Conclusions (Sec.~\ref{sec:summary}).

%%%%%%%%%%%%%%%%%%%%%%%%%%%%%%%%%%%%%%%%%%%%%%%%%%%%%%%%%%%%%%%%%%%%%%%%%%%%%%%%%%%%%%%%%%%%%%%%
\section{Model \label{sec:model}}

\begin{figure}[H]
    \centering
    \includegraphics[width=0.7\columnwidth]{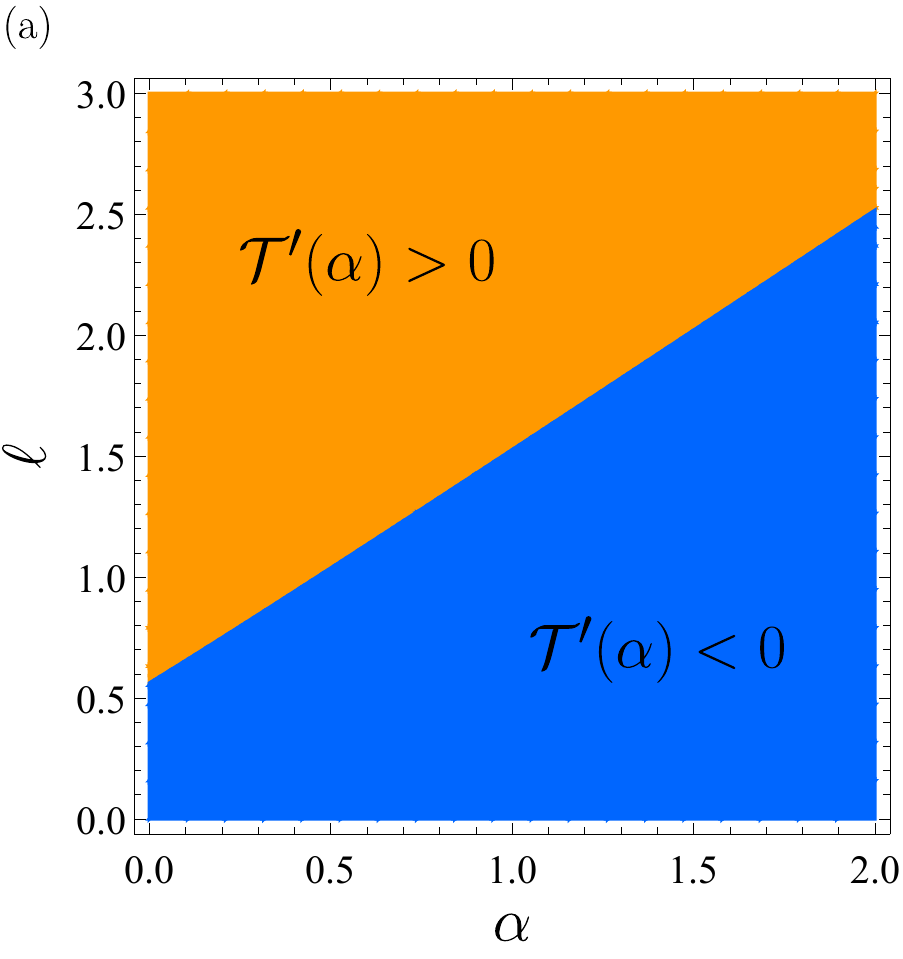}\\
    \includegraphics[width=0.8\columnwidth]{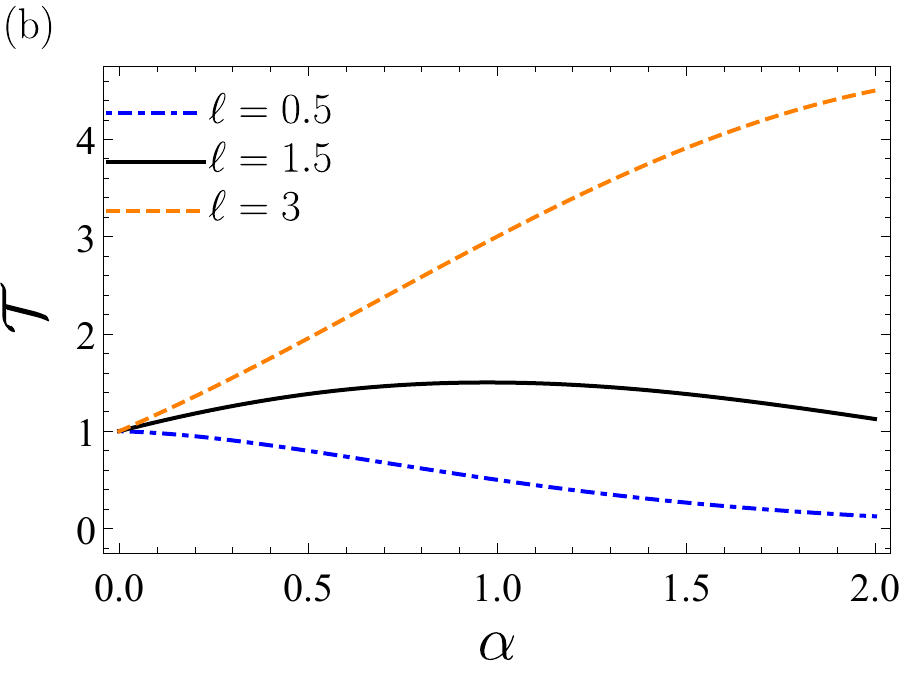}\\
    \caption{    Phase diagram (top panel -- $(a)$) showing domains where MFPT is the increasing ($\mathcal{T}'(\alpha)>0$) (orange) or decreasing ($\mathcal{T}'(\alpha)<0$) (blue) function of the stability index $\alpha$ and
    sample dependence of $\mathcal{T}(\alpha)$ (bottom panel -- $(b)$) corresponding to all (three) possible shapes of MFPT curves ($\ell=0.5$ --- blue dash-dotted,  $\ell=1.5$ --- black solid and \kct{$\ell=3$} --- orange dashed).}
\label{fig:phasediagram}
\end{figure}

We consider the overdamped motion described by the following Langevin \cite{gardiner2009} equation
\begin{equation}
    \dot{x}(t)=\zeta (t),
    \label{eq:langevin}
\end{equation}
where $\zeta(t)$ is a symmetric $\alpha$-stable noise \cite{janicki1994b,dubkov2008}.
The $\alpha$-stable noise is the formal time derivative of the $\alpha$-stable process $L(t)$, see Ref.~\cite{janicki1994b}, which is the process with stationary, independent increments distributed according to the $\alpha$-stable distribution \cite{samorodnitsky1994}.
Values of the symmetric $\alpha$-stable motion $L(t)$ are distributed according to the symmetric $\alpha$-stable distribution which is defined by the characteristic function
\begin{equation}
 \phi(k)=\left\langle \exp[ik {L}(t)] \right\rangle=\exp\left[ -  t \sigma^\alpha |k|^\alpha \right],
 \label{eq:fcharakt}
\end{equation}
where $\alpha$ ($0<\alpha \leqslant 2$) is the stability index, while $\sigma$ ($\sigma>0$) is the scale parameter.
The stability index $\alpha$ determines the tails' asymptotics, which for $\alpha<2$ is of the power-law type, i.e., $p(x) \propto |x|^{-(\alpha+1)}$.
Furthermore, the noise driven motion, see Eq.~(\ref{eq:langevin}), is restricted by two absorbing boundaries placed at $\pm l$, i.e., the motion is performed within a bounded $[-l,l]$ domain.
For the system described by Eq.~(\ref{eq:langevin}), it is possible to define the  mean first passage time (MFPT) $\mathcal{T}$
\begin{eqnarray}
    \mathcal{T}  =  \langle \tau \rangle =
     \langle \min\{\tau : x(0)=0 \;\land\; |x(\tau)| \geqslant l  \} \rangle.
    \label{eq:mfpt-def}
\end{eqnarray}
The exact formula for the MFPT \cite{getoor1961,blumenthal1961} in the setup described above reads
\begin{equation}
\mathcal{T}=\frac{1}{\Gamma(1+\alpha)} \ell^{\alpha},
    \label{eq:mfpt}
\end{equation}
where $\ell=l/\sigma$.
Using Eq.~(\ref{eq:mfpt}), it is possible to draw a phase-diagram showing domains where $\mathcal{T}(\alpha)$ is decreasing ($\mathcal{T}'(\alpha)<0$ --- colored in blue) or increasing ($\mathcal{T}'(\alpha)>0$ --- colored in orange) function of the stability index $\alpha$, see top panel of Fig.~\ref{fig:phasediagram}.
For $\ell \geqslant \exp(\frac{3}{2}-\gamma) \approx 2.51$, where $\gamma$ is the Euler-Mascheroni constant, the mean first passage time increases with the increase of $\alpha$, while for
$\ell \leqslant \exp(-\gamma) \approx 0.56$ the MFPT is the decreasing function of $\alpha$.
In the intermediate domain, $ \exp(-\gamma) < \ell < \exp(\frac{3}{2}-\gamma)$, the mean first passage time is a non-monotonous function of the stability index $\alpha$.
Moreover, there exists such a value of the stability index $\alpha$, let say $\alpha_c$ ($0<\alpha_c<2$), for which the MFPT attains maximum value.
The bottom panel of Fig.~\ref{fig:phasediagram} shows all (three) possible patterns of MFPT curves.
In particular, the blue dash-dotted line shows results for  $\ell \leqslant \exp(-\gamma)$, i.e., $\ell=0.5$, the black solid line depicts results for $\exp(-\gamma) < \ell < \exp(3/2-\gamma)$, i.e., $\ell=1.5$, and orange dashed for $\ell \geqslant \exp(3/2-\gamma)$, i.e., \kct{$\ell=3$}.
These three lines present: decreasing, non-monotonous and increasing dependence of the mean first passage time on the stability index $\alpha$.
In the limit of $\alpha=0$, the MFPT is independent of $\ell$ making all curves starting at the same point, see Eq.~(\ref{eq:mfpt}).

\bdt{
The existence of the domain of non-monotonous dependence of MFPT on the stability index $\alpha$ can be intuitively explained.
In order to escape from a narrow interval, i.e., small $\ell$, a sequence of short jumps is sufficient.
Short jumps are controlled by the central part of the jump length distribution, which is a growing function of $\alpha$.
Therefore, the smallest MFPT is recorded for $\alpha=2$.
The very different situation is observed for large $\ell$.
In such a case, the most probable escape scenario is via a single long jump \cite{imkeller2006,imkeller2006b}.
Consequently, densities with heavier tails result in a faster escape and the minimal MFPT is recorded for $\alpha=0$.
Finally, there is an intermediate domain of $\ell$ where the competition between short and long jumps is observed in which MFPT is a non-monotonous function of $\alpha$.
}

Here, instead of constant value of the stability index $\alpha$, we assume that the parameter $\alpha$ periodically changes in time
\begin{equation}
    \alpha(t) =\overline{\alpha}+ \frac{\Delta \alpha}{2} \sin{\left(2\pi f t\right)},
    \label{eq:alphaChange}
\end{equation}
where
\begin{equation}
\overline{\alpha}=\frac{\amin+\amax}{2}
\end{equation}
and
\begin{equation}
\Delta \alpha = \amax-\amin.
\end{equation}
For the modulation given by Eq.~(\ref{eq:alphaChange}) the mean value of the stability index $\alpha$ over the period of modulation, $T$, is equal to $\overline{\alpha}$.
\bdt{Despite the fact that, by construction, $\Delta \alpha>0$ we use the $\Delta \alpha$ with positive or negative signs in order to indicate the initial dependence of $\alpha$ on time.
Therefore, for  $\Delta \alpha>0$ we have $\dot{\alpha}(0)>0$ while for  $\Delta \alpha<0$ there is $\dot{\alpha}(0)<0$.
}
In further studies, we are using $\Delta \alpha = \pm 1$.
Consequently, for $\Delta \alpha=1$, $\alpha(t)$ initially increases, while for $\Delta \alpha=-1$ it decays.
Despite the fact that  $\alpha$ is no longer constant, Eq.~(\ref{eq:mfpt}) provides a qualitative explanation of dependence of the MFPT on the frequency $f$.
From Eq.~(\ref{eq:mfpt}) it implies whether MFPT is increasing, decreasing or non-monotonous function of the stability index $\alpha$.
The smaller MFPT corresponds to the situation when the first passage time density is narrower, because the asymptotics of the first passage time density can be approximated by $p(\tau|\alpha) \sim \exp(-\frac{\kct{\tau}}{\mathcal{T}(\alpha)})$ \cite{dybiec2017levy}.
Putting it differently, the width of the instantaneous first passage time density has qualitatively the same dependence on the stability index $\alpha$ as the mean first passage time $\mathcal{T}(\alpha)$, see Eq.~(\ref{eq:mfpt}).
When first passage time density is narrower individual escapes are (statistically) faster.
If individual escapes, for a given value of $\alpha$, become faster the escape kinetics is facilitated.
Therefore, in the time dependent case, $\alpha$ with smaller $\mathcal{T}(\alpha)$, on average, speeds up escapes, while $\alpha$ with the larger MFPT statistically slows down the escape kinetics.
The overall efficiency of the escape kinetics is determined by the time scale associated with the modulation of $\alpha$ and the initial direction of changes in the value of the stability index.
In the next section we present numerical results for MFPTs with periodically modulated stability index $\alpha$.

%%%%%%%%%%%%%%%%%%%%%%%%%%%%%%%%%%%%%%%%%%%%%%%%%%%%%%%%%%%%%%%%%%%%%%%%%%%%%%%%%%%%%%%%%%%%%%%%
% \clearpage
\section{Results \label{sec:results}}

\begin{figure}[H]
    \centering
    \includegraphics[width=0.9\columnwidth]{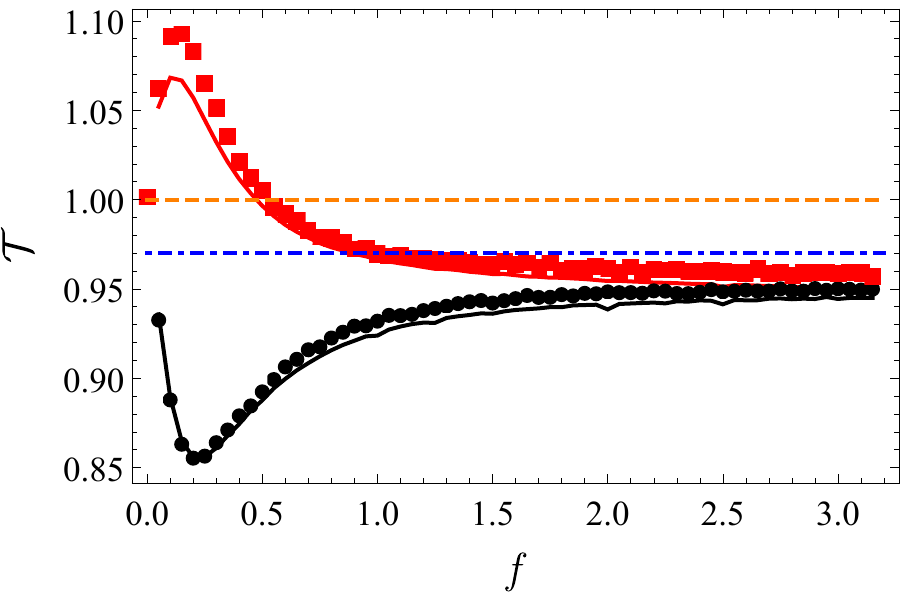}\\
    \caption{The dependence of the MFPT on the driving frequency $f$. Points represent results of computer simulations with $\Delta \alpha=1$ (black dots)  and $\Delta \alpha=-1$ (red squares), $\alpha \in [0.5,1.5]$ and $\ell=1$.
    The orange dashed line represents $\mathcal{T}(\overline{\alpha})$, while the blue dash-dotted line shows $\langle\mathcal{T}(\alpha)\rangle_{\mathrm{arcsine}}$, see Eq.~(\ref{eq:arcsine}).
    Finally, red and black solid curves show $\langle \mathcal{T}(\alpha) \rangle_{p(\alpha)}=\int_{\amin}^{\amax}p(\alpha) \mathcal{T}(\alpha) d\alpha$, where exemplary $p(\alpha)$ are depicted in Fig.~\ref{fig:alfaWyjsciaAlfa1L1}
    Error bars, representing the standard deviation of the mean, are within the symbol size.
    }
\label{fig:mfptOscAlfa1l1}
\end{figure}

\begin{figure}[H]
    \centering
    \includegraphics[width=0.95\columnwidth]{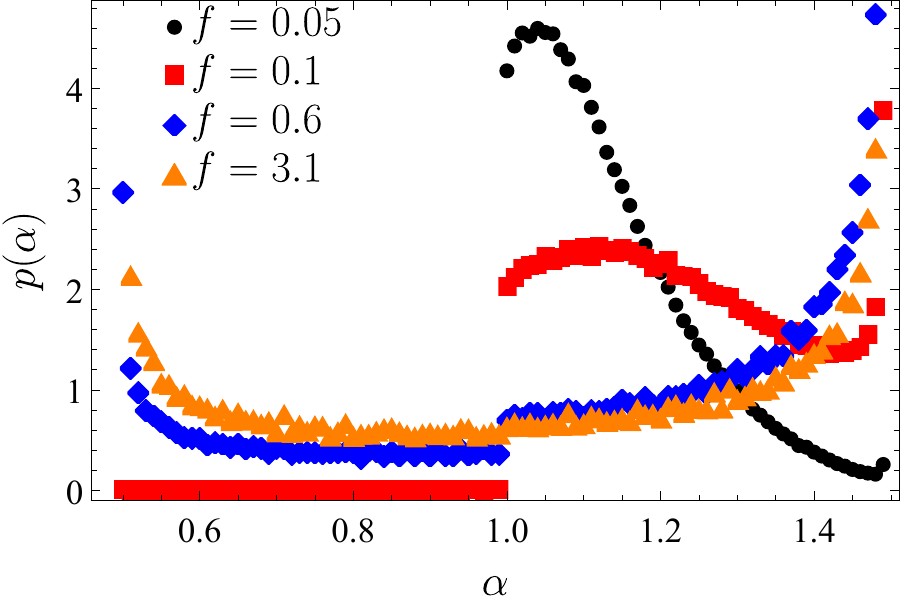}
    \caption{Histograms of instantaneous values of the stability index $\alpha$, $p(\alpha)$, at first passage time for $\alpha\in [0.5,1.5]$ with $\ell=1$.
    Different curves correspond to various values of the  driving frequency $f$ (black dots, red squares, blue and orange triangles correspond to $f\in\{0.05,0.1,0.6,3.1\}$ respectively).}
    \label{fig:alfaWyjsciaAlfa1L1}
\end{figure}

\begin{figure}[H]
    \centering
    \includegraphics[width=0.9\columnwidth]{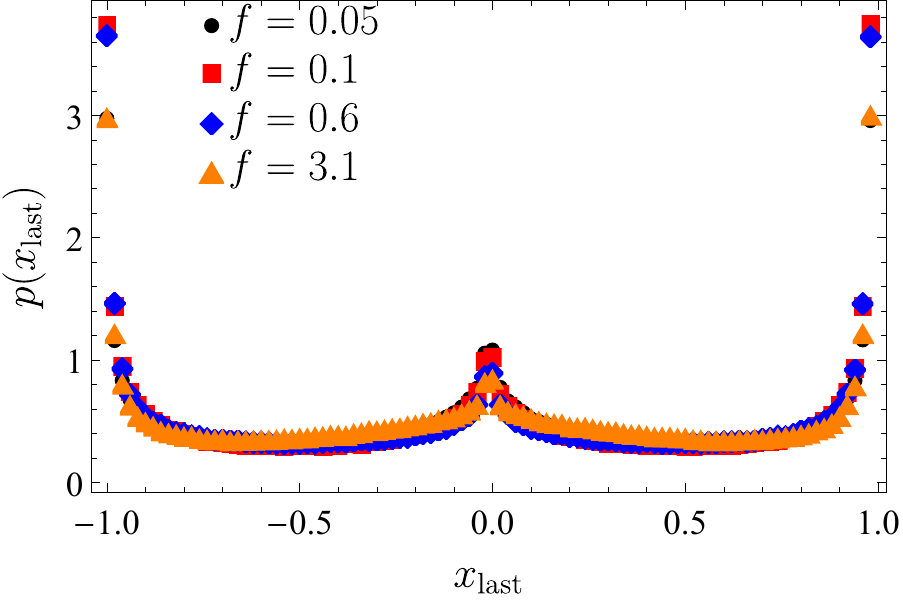}
    \caption{Histograms of last hitting points, $p(x_{\mathrm{last}})$  for $\alpha\in [0.5,1.5]$ with $\ell=1$.
    Different curves correspond to various values of the frequency $f$ (black dots, red squares, blue and orange triangles correspond to $f\in\{0.05,0.1,0.6,3.2\}$ respectively).}
    \label{fig:last}
\end{figure}

\begin{figure}[H]
    \centering
    \includegraphics[width=0.9\columnwidth]{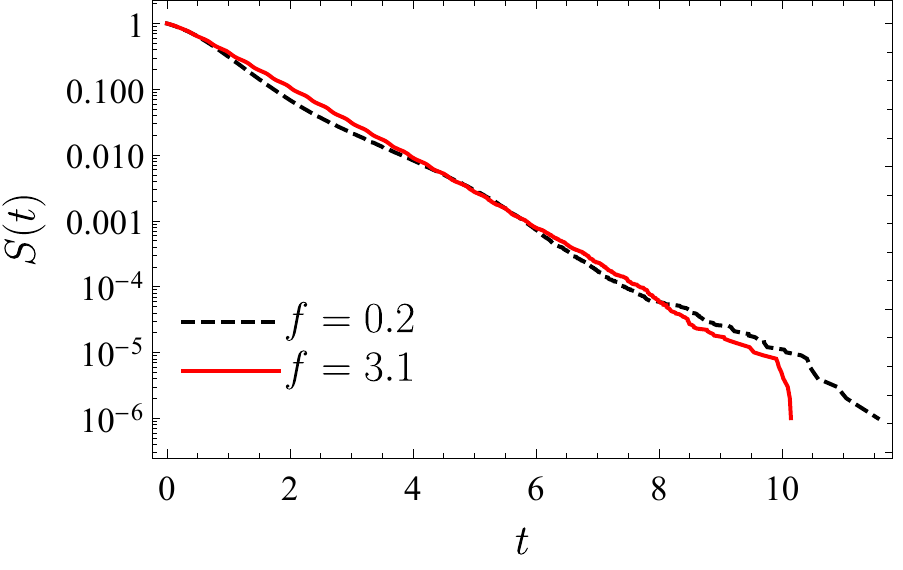}
    \caption{Survival probability $S(t)$ for $\alpha\in [0.5,\kct{1.5}]$, $\ell=1$ with $\Delta \alpha=1$.
    Various curves correspond to different values of $f$: \kct{$f=0.2$} (black dashed) and $f=3.1$ (red solid) .}
    \label{fig:survival}
\end{figure}

\begin{figure}[H]
    \centering
    \includegraphics[width=0.9\columnwidth]{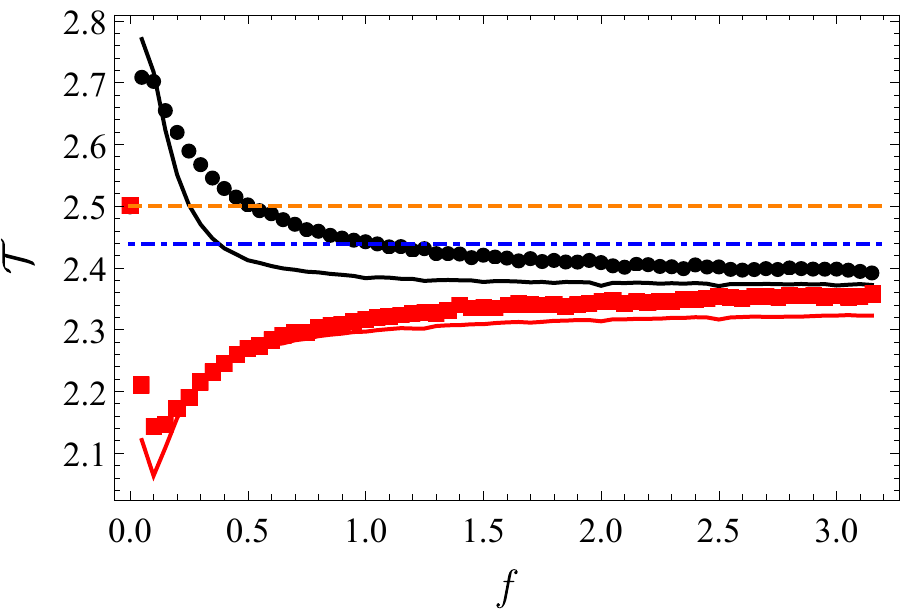}\\
    \caption{The same as in Fig.~\ref{fig:mfptOscAlfa1l1} for $\ell=2.5$.}
    \label{fig:mfptOscAlfa1l2}
\end{figure}

\begin{figure}[H]
    \centering
    \includegraphics[width=0.9\columnwidth]{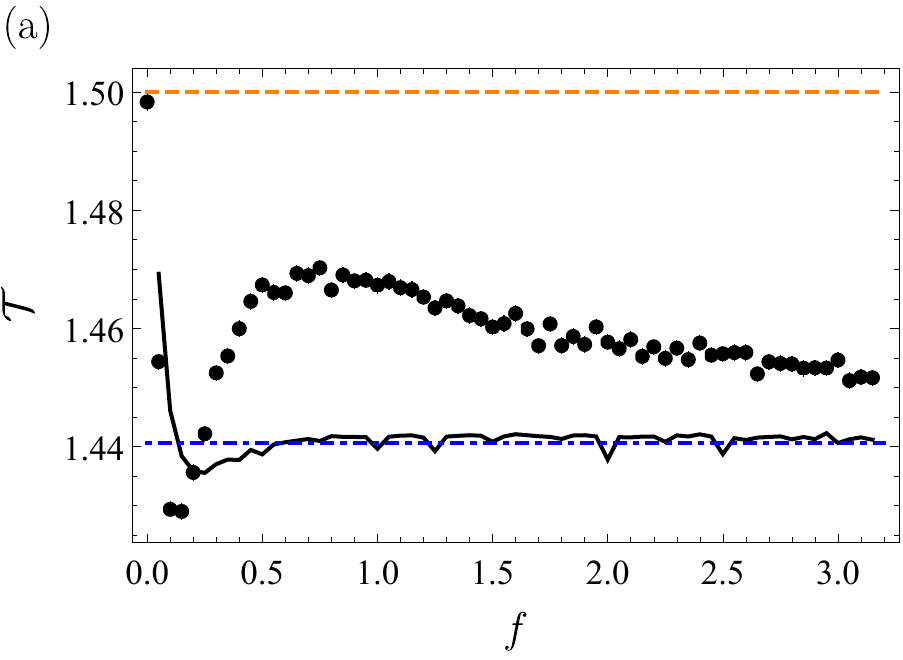}\\
    \includegraphics[width=0.9\columnwidth]{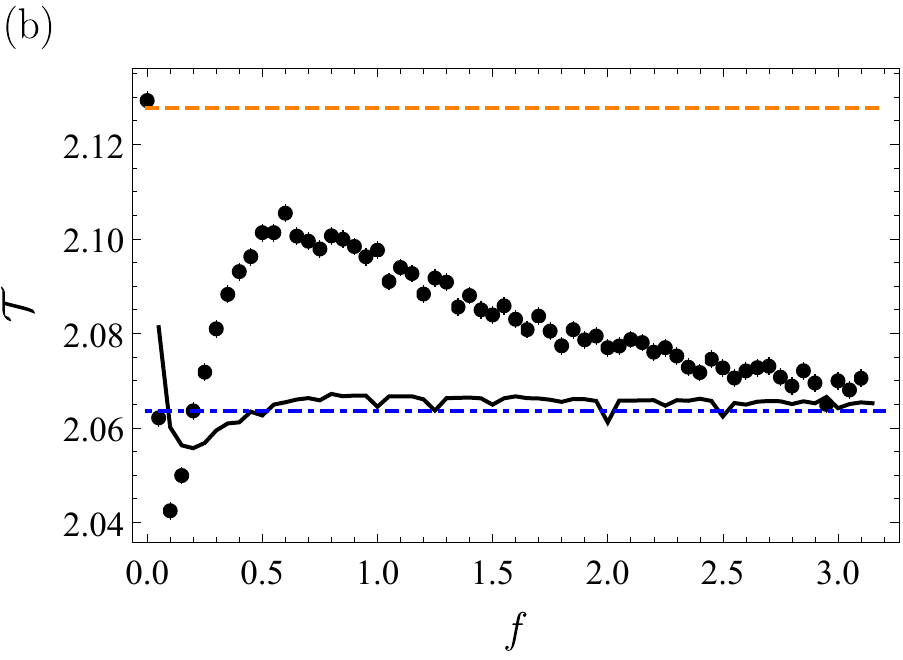}
    \caption{The same as in Fig~\ref{fig:mfptOscAlfa1l1} for $\Delta \alpha=1$ with $[\amin,\amax]=[0.5,1.5]$ and $\ell=1.5$ (top panel -- $(a)$) and  $[\amin,\amax]=[1,2]$ and $\ell=2$ (bottom panel -- $(b)$).
    $\overline{\alpha}=(\amin+\amax)/2$ and $\ell$ are chosen in such a way that $\mathcal{T}'(\alpha)$ changes its sign during the modulation of $\alpha$.
    Error bars represent the standard deviation of the mean.
}
    \label{fig:mfptNearCritical}
\end{figure}

Using the Euler-Maruyama method \cite{janicki1994,janicki1996}, we have generated an ensemble of trajectories  following Eq.~(\ref{eq:langevin}).
Every trajectory was simulated until the first escape from the $[-l,l]$ interval, i.e., as long as $|x(t)| < l$.
From the set of collected first passage times $\tau$s the mean first passage time $\mathcal{T}=\langle \tau \rangle$ was calculated, see Eq.~(\ref{eq:mfpt-def}).
Obtained results are presented in a series of figures showing MFPTs and other motion characteristics (Figs.~\ref{fig:mfptOscAlfa1l1} -- \ref{fig:survival}), while further Figs.~\ref{fig:mfptOscAlfa1l2} and~\ref{fig:mfptNearCritical} show MFPTs for other sets of parameters.
The studied system is characterized by two time scales.
The first time scale is determined by the periodic modulation of the stability index $\alpha$, which is characterized by its period $T$ ($T=1/f$).
The second time scale is imposed by the escape kinetics and it is determined by the MFPT.
Due to modulation of $\alpha$, the stability index $\alpha$ is no longer constant during the motion.
Different escapes are recorded at various values of instantaneous $\alpha$.
In the dynamic regime, recorded mean first passage times  are always between the minimum and maximum of mean first passage times with fixed $\alpha$, i.e.,
$\min_{\alpha\in [\amin,\amax]}\{\mathcal{T}(\alpha)\} \leqslant \mathcal{T} \leqslant \max_{\alpha\in [\amin,\amax]}\{\mathcal{T}(\alpha)\}$.

We start our analysis with such values of $\overline{\alpha}$, $\Delta \alpha$ and $\ell$ that $\mathcal{T}'(\alpha)$ does not change its sign with the change in the stability index $\alpha$.
In other words, $\mathcal{T}'(\alpha)$ is always smaller or larger than 0.
In Fig.~\ref{fig:alfaWyjsciaAlfa1L1}  results corresponding to $\mathcal{T'(\alpha})<0$ are depicted, see blue domain in top panel of Fig.~\ref{fig:phasediagram}.
We have used  $\alpha \in [\amin,\amax]=[0.5,1.5]$, $\Delta \alpha=\pm 1$ and $\ell=1$ resulting in $\overline{\alpha}=1$ and $\mathcal{T}'(\alpha) < 0$ for every instantaneous value of the stability index $\alpha$.
Points depict results of computer simulations with $\Delta \alpha=1$ (black dots) and $\Delta \alpha=-1$ (red squares).
Lines in Fig.~\ref{fig:mfptOscAlfa1l1} show $\mathcal{T}(\overline{\alpha})$ (orange dashed) and $\langle\mathcal{T}(\alpha)\rangle_{\mathrm{arcsine}}$ (blue dash-dotted), see below.
The recorded dependence of the MFPT on the frequency $f$ follows a resonant-activation-like \cite{doering1992} ($\Delta \alpha=1$) or noise-enhanced-stability-like \cite{agudov2001} ($\Delta \alpha=-1$) patterns.
More precisely, for $\Delta \alpha=1$, $\alpha(t)$ initially grows.
In subsequent moments $\alpha$ becomes larger and $\mathcal{T}(\alpha)$ smaller, as we are in the $\mathcal{T}'(\alpha)<0$ domain.
For $t<T/2$, with the increasing time chances of escape increases.
As long as $\mathcal{T} \ll T$, the modulation of $\alpha$ facilitates the escape kinetics.
Therefore, there exists such a value of $f$ for which the MFPT attains, analogously like in the resonant activation, the minimal value.
Further increase in $f$ makes the MFPT larger.
For $\Delta \alpha=-1$, in comparison to $\Delta \alpha=1$, the MFPT curve is inverted because initial decay in $\alpha$ is associated with the decreasing chances of escape.
Therefore, the MFPT curve displays the noise-enhanced-stability-like behavior: there exists such a value of the frequency $f_c$ for which the MFPT is maximal.
\bdt{In the case of $\Delta \alpha=-1$, the pattern of MFPT curve is analogous to a non-monotonous behavior of the MFPT as a function of the modulation frequency in the presence of a metastable potential \cite{agudov2001}.
For the escape from metastable potential the MFPT is sensitive to the barrier configuration and depends on parameters characterizing a modulation protocol.
In our setup, there is no deterministic force but the MFPT depends on the stability index $\alpha$ and its variation.
}
The MFPT curve has a single extreme if changes in $\alpha$ do not change the sign of $\mathcal{T}'(\alpha)$.
For $f=0$, the MFPT is equal to $\mathcal{T}(\alpha(0))=\mathcal{T}(\overline{\alpha})$, see Eq.~(\ref{eq:mfpt}), because the stability index $\alpha$ is constant and equal to its initial value.
For $f_c$ the minimum (if $\sign (\Delta \alpha)= -\sign(\mathcal{T}'(\alpha))$) or the maximum (if $\sign (\Delta \alpha)= \sign(\mathcal{T}'(\alpha))$) is recorded.

The main quantity which characterizes the escape kinetics is the mean first passage time.
Furthermore, the escape kinetics can be characterized by the instantaneous value of the stability index $\alpha$ at the moment of first escape, i.e., at the first passage time.
% \karol{Dla mnie dwa kolejne zdania to w pewien sposób powtarzanie tego samego.}
Fig.~\ref{fig:alfaWyjsciaAlfa1L1} shows histograms $p(\alpha)$ for various values of the frequency $f$ corresponding to MFPTs depicted in Fig.~\ref{fig:mfptOscAlfa1l1}.
% In Fig.~\ref{fig:alfaWyjsciaAlfa1L1} the same parameters as in Fig.~\ref{fig:mfptOscAlfa1l1} are used ($\amin=0.5$, $\amax=1.5$, $\ell=1$ and $\Delta \alpha=1$), i.e., the system is in $\mathcal{T}'(\alpha) < 0$ domain.
For $f=0$, the $p(\alpha)$ density is given by the Dirac's delta ($p(\alpha)=\delta(\alpha-\overline{\alpha})$), because all escapes take place with $\alpha=\overline{\alpha}$.
For $f=0.05$, there is a single maximum near the $\alpha=\overline{\alpha}$ and slowly decaying part towards $\amax=1.5$.
With increasing $f$ the height of the central peak at $\alpha=\overline{\alpha}$ decreases and the maximum at $\alpha=\amax$ emerges.
For $f=0.1$ the central peak does not decay completely however most particles exit with $\alpha \approx \amax$ as it corresponds to the minimal MFPT.
For small $f$ almost all escapes take place during the time when $\alpha$ has not managed to drop down below $\overline{\alpha}$.
Therefore, for $\alpha < \overline{\alpha}$, the histogram vanishes, i.e., $p(\alpha) \equiv 0$.
With the further increase in $f$ the non-zero probability $p(\alpha)$ for $\alpha<\overline{\alpha}$ emerges, because a substantial fraction of escape events is recorded for small values of the stability index $\alpha$.
For $f=0.6$ and $f=3.1$ most particles escape with extreme values of $\alpha$, i.e. $\alpha \approx \amin$ or $\alpha \approx \amax$, however for $f=0.6$ the $p(\alpha)$ still has the discontinuity at $\alpha=\overline{\alpha}$.
Finally, for large $f$, e.g., $f=3.1$, the $p(\alpha)$ density is almost symmetric along $\alpha=1$.
Nevertheless, for $f$ which is not large enough $p(\alpha)$ densities are skewed into the direction of $\Delta \alpha$.
The change in $\Delta \alpha$ from $\Delta \alpha=1$ to $\Delta \alpha=-1$ reflects the $p(\alpha)$ density along the $\alpha=\overline{\alpha}=1$ line.

The $p(\alpha)$ distribution with $f\to\infty$ approaches
\begin{equation}
p_{\infty}(\alpha)= \frac{2 }{ \pi \sqrt{\Delta \alpha^2-4(\alpha-\overline{\alpha})^2}},
    \label{eq:rozkladyAlfa}
\end{equation}
which is of the analogous shape like $p(v)$ and $p(x)$ distributions in the L\'evy walk scenario in the parabolic potential \cite{dybiec2018conservative}.
The density given by Eq.~(\ref{eq:rozkladyAlfa}) is of the arcsine type and it is normalized on $\alpha \in [\amin,\amax]=[\overline{\alpha}-\Delta\alpha/2,\overline{\alpha}+\Delta\alpha/2]$.
If one knows the first passage time density $p(\tau)$, using transformation of variables, it is possible to obtain the $p(\alpha)$ distribution.
From numerical simulation (results not shown), we see that for $f$ large enough $\tau\;\mathrm{mod}\;T$ is approximately uniform on the $[0,T)$ interval.
Consequently, the argument of $\sin$ in Eq.~(\ref{eq:alphaChange}) is uniformly distributed over the $[0,2\pi)$ interval and $p(\alpha)$ approaches the arcsine distribution $p_\infty(\alpha)$, see Eq.~(\ref{eq:rozkladyAlfa}).
From Eq.~(\ref{eq:rozkladyAlfa}) it is possible to calculate $\langle\mathcal{T}(\alpha) \rangle_{\mathrm{arcsine}}$, i.e.,
\begin{equation}
\langle\mathcal{T}(\alpha) \rangle_{\mathrm{arcsine}} = \int_{\amin}^{\amax} p_\infty(\alpha) \mathcal{T}(\alpha) d\alpha,    
\label{eq:arcsine}
\end{equation}
which is marked with a blue dash-dotted line in Fig.~\ref{fig:mfptOscAlfa1l1}.
Moreover, using the distribution $p(\alpha)$ at the first passage time one can calculate
\begin{eqnarray}
    \langle \mathcal{T}(\alpha) \rangle_{p(\alpha)} = \int_{\amin}^{\amax} p(\alpha) \mathcal{T}(\alpha) d\alpha.
    \label{eq:integral}
\end{eqnarray}

In the limit of $f \to \infty$ one could expect that $\mathcal{T} \to \mathcal{T}(\langle \alpha \rangle)$, but actually one sees that $\mathcal{T}$ is closer to $\langle \mathcal{T}(\alpha ) \rangle_{\mathrm{arcsine}}$, see Eqs.~(\ref{eq:rozkladyAlfa}) and~(\ref{eq:arcsine}).
Therefore, asymptotic properties of escape from finite intervals induced by the $\alpha$-stable noise with the time dependent stability index are very different from asymptotic properties of resonant activation \cite{pankratov2000}.
In \cite{pankratov2000} it has been shown that in the $f\to\infty$ limit the mean first passage time over periodically modulated potential barrier is equal to the MFPT over the potential averaged over the period of modulation.
Here, it is the other way round as $\alpha$ averaged over $p_\infty(\alpha)$ is equal to $\overline{\alpha}$, which gives the $f=0$ limit.
Such an approximate limiting behavior for $f\to\infty$ arises due to general properties of escapes induced by the $\alpha$-stable noise.
Under the $\alpha$-stable driving, the most probable escape scenario is the escape via a single long jump.
As it can be seen from Fig.~\ref{fig:last}, a significant fraction of particles waits for an extreme jump and then it escapes with a fixed, instantaneous, value of the stability index $\alpha$.
Therefore, a sequence of escapes with various values of $\alpha$ is recorded.
The level of agreement  between $\mathcal{T}$ and $\langle \mathcal{T}(\alpha) \rangle_{\mathrm{arcsine}}$ depends on the system parameters.
Interestingly, using the numerically estimated $p(\alpha)$ distribution it is possible to calculate $\langle \mathcal{T}(\alpha) \rangle_{p(\alpha)} $.
In Fig.~\ref{fig:mfptOscAlfa1l1}, these estimates are plotted with red solid ($\Delta\alpha=-1$) and black solid ($\Delta\alpha=1$) curves.
They nicely follow results of computer simulations (points).
The agreement is most likely coincidental, because $\langle \mathcal{T}(\alpha) \rangle_{p(\alpha)}$ corresponds to evolution with the fixed $\alpha$ distributed according to $p(\alpha)$, similarly like for distributed order fractional derivatives \cite{sokolov2004distributed,meerschaert2011distributed}, while here $\alpha$ changes deterministically in time.

The distribution of the stability index $\alpha$ at the first passage time $p(\alpha)$ can be contrasted with the distribution $p_{t} (\alpha)$ of instantaneous values of the stability index $\alpha$ prior to the escape
$p_t(\alpha)     = \langle  \delta(\alpha-\alpha(t))\rangle$.
If first passage time is long enough such a distribution also tends to the arcsine distribution given by Eq.~(\ref{eq:rozkladyAlfa}), but this time the convergence rate is much faster because every trajectory adds a whole ensemble of $\alpha$s.
Therefore, contrary to $p(\alpha)$ distribution, the arcsine distribution can be recorded for finite frequencies also.
For instance, for the setup studied in Fig.~\ref{fig:mfptOscAlfa1l1}, already for $f>0.2$, $p_t(\alpha)$ distribution is indistinguishable from the arcsine distribution, see Eq.~(\ref{eq:rozkladyAlfa}).
From Fig.~\ref{fig:alfaWyjsciaAlfa1L1} it is clearly visible that the value of the stability index $\alpha$ at the escape time follow a different pattern than the during-the-motion distribution of instantaneous values of the stability index.
Nevertheless, in both cases, i.e., for $p(\alpha)$ and $p_t(\alpha)$, the same limiting density is reached in the $f\to\infty$ limit.
One can conclude, that the instantaneous value of the stability index $\alpha$ at the first escape does not need to be the most probable value of the stability index which is recorded during the motion.

Fig.~\ref{fig:last} complements examination of the properties of escape scenarios performed in Fig.~\ref{fig:mfptOscAlfa1l1}.
It shows $p(x_{\mathrm{last}})$, i.e., histograms of last visited points ($x_{\mathrm{last}}$) before leaving the $[-1,1]$ interval.
It clearly indicates that a significant fraction of particles escape from the initial point, i.e., from $x_{\mathrm{last}}=0$, while majority of particles approach absorbing boundaries at $\pm l$.
The $p(x_{\mathrm{last}})$ distribution is symmetric along $x_{\mathrm{last}}=0$ reflecting symmetry of the noise.

Fig.~\ref{fig:survival} finishes exploration of the setup studied in Fig.~\ref{fig:mfptOscAlfa1l1} ($\alpha\in[0.5,1.5]$, $\ell=1$ and $\Delta \alpha=1$).
It presents the survival probability $S(t)$, i.e., the probability that at time $t$ a particle is still in the $[-l,l]$ interval, for $f=0.2$ (black dashed line) and $f=3.1$ (red solid line).
The survival probability is a decaying function of time because with the increasing time chances of finding a particle in the domain of motion decay.
Furthermore, the survival probability displays a typical exponential trend which can be decorated by some bending due to modulation in $\alpha$.
Such a bending is especially well visible for small values of frequencies, e.g., $f=0.2$ (black dashed line).

In Fig.~\ref{fig:mfptOscAlfa1l2}  the \bdt{rescaled} half-width of the interval is set to $\ell=2.5$ consequently for all recorded values of $\alpha$ ($\alpha \in [0.5,1.5]$) $\mathcal{T'(\alpha})>0$, see the orange domain in top panel of Fig.~\ref{fig:phasediagram}.
Therefore, in comparison to $\ell=1$, the increase in $\ell$ from 1 to 2.5 exchanges monotonicity of MFPT curves, cf. Fig.~\ref{fig:mfptOscAlfa1l1} and Fig.~\ref{fig:mfptOscAlfa1l2}.
Dashed line in Fig.~\ref{fig:mfptOscAlfa1l2} shows $\mathcal{T}(\overline{\alpha})$ while the blue dash-dotted line $\langle\mathcal{T}(\alpha)\rangle_{\mathrm{arcsine}}$.
Additional solid black and red curves show $ \langle \mathcal{T}(\alpha) \rangle_{p(\alpha)} $.
This time the level of agreement between results of computer simulations and $\langle \mathcal{T}(\alpha) \rangle_{p(\alpha)}$ approximation is worse than in \kct{Fig.~\ref{fig:mfptOscAlfa1l1}}.
The decrease of agreement is produced by the dynamics prior to the last escape.
Increase in $\ell$ from 1 to 2.5 increases the mean first passage time $2.5^\alpha$ times.
Consequently, the escape process is slower and particles have more time to diffuse making integration of Eq.~(\ref{eq:mfpt}) over $p(\alpha)$ not fully reliable.
For $\Delta \alpha=1$, in the limit of $f\to\infty$ approaches $\langle\mathcal{T}(\alpha)\rangle_{\mathrm{arcsine}}$.

Finally, in Fig.~\ref{fig:mfptNearCritical}, the model parameters are adjusted in such a way that $\mathcal{T}'(\alpha)$ changes its sign during periodic modulation of $\alpha$.
We have used two sets of parameters: (i) $\ell=1.5$, $\amin=0.5$, $\amax=1.5$ with $\alpha(0)=1$ (top panel) and (ii) $\ell=2$, $\amin=1.0$, $\amax=2.0$ with $\alpha(0)=1.5$ (bottom panel).
In such a case in (i) $\mathcal{T}'(0.973) \approx 0$ while in (ii) $\mathcal{T}'(1.479) \approx 0$ and $\alpha(0)$ in both cases lies in the domain where $\mathcal{T}'(\kct{\alpha(0)})<0$.
Therefore, for $\Delta \alpha=1$, with the increasing $\alpha$ the $\mathcal{T}'(\alpha)$ changes its sign from negative to positive.
Such an initial condition allows for the rapid decay of the MFPT with very slowly increasing $\alpha$, i.e. for small $f$.
Further increase in the frequency is sufficient to move $\alpha$ to the value for which $\mathcal{T}'(\alpha)$ becomes positive, which in turn increases the MFPT.
Therefore, in the situation when $\mathcal{T}'(\alpha)$ changes its sign, in addition to the maximum of MFPT, there is a local, narrow minimum at a  small $f$ ($f \approx 0.1$).
The change of sign in $\Delta \alpha$ from $\Delta \alpha=1$ to $\Delta \alpha=-1$, analogously like in Figs.~\ref{fig:mfptOscAlfa1l1} and~\ref{fig:mfptOscAlfa1l2}, inverts the shape of MFPT curves, i.e., there is a narrow maximum at small $f$ and wide minimum at larger $f$.
Finally in the limit of $f\to\infty$ $\mathcal{T}$ approaches $\langle \mathcal{T}(\alpha) \rangle_{\mathrm{arcsine}}$ but for $\ell=2$ (bottom panel of Fig.~\ref{fig:mfptNearCritical}) agreement is better.
Furthermore, as it can be seen from Fig.~\ref{fig:mfptNearCritical}, the mean first passage time estimated using Eq.~(\ref{eq:integral}) significantly deviates from numerically obtained values of MFPTs.
This discrepancy originates in the fact, that for the setup studied in Fig.~\ref{fig:mfptNearCritical}, the $p(\alpha)$ density attains the arcsine $p_\infty(\alpha)$ shape already for finite $f$.

%%%%%%%%%%%%%%%%%%%%%%%%%%%%%%%%%%%%%%%%%%%%%%%%%%%%%%%%%%%%%%%%%%%%%%%%%%%%%%%%%%%%%%%%%%%%%%%%
% \clearpage
\section{Summary and conclusions \label{sec:summary}}

The resonant activation is one of phenomena manifesting constructive role of fluctuations.
The resonant activation is a generic effect for barrier crossing events in the conformally modulated energy landscape \cite{doering1992,pankratov2000}, i.e. for a given rate of periodic or dichotomous modulation of the potential barrier the minimal average escape
time is observed.

Here, we have studied properties of the minimal setup allowing for occurrence of the resonant activation.
In comparison to the typical models we have reduced the number of elements by eliminating the deterministic force.
The model of L\'evy noise induced escape from finite intervals is capable of revealing the phenomenon of the resonant activation if the stability parameter of the noise is periodically modulated.
Consequently, like in the resonant activation phenomenon, there is such a value of modulating frequency, for which the mean first passage time is minimal.
The model can reveal not only the resonant activation but also the noise enhanced stability, because the mean first passage time can be not only decreased but also increased.
The model itself displays sensitivity to the initial direction in modulation of the stability index $\alpha$, as it inverts the shape of mean first passage time curves.
At the same time, survival probabilities follow exponential decay and distribution of instantaneous values of stability index $\alpha$ at first passage time are skewed. The direction of asymmetry is determined by initial monotonicity of modulation.
Finally, non trivial asymptotic behavior is recorded, i.e., in the high frequency limit recorded mean first passage time does  not correspond to mean first passage time with average value of the stability index $\alpha$.

%%%%%%%%%%%%%%%%%%%%%%%%%%%%%%%%%%%%%%%%%%%%%%%%%%%%%%%%%%%%%%%%%%%%%%%%%%%%%%%%%%%%%%%%%%%%%%%%
\section*{Acknowledgements}

This research was supported in part by PLGrid Infrastructure and by the National Science Center (Poland) grant 2018/31/N/ST2/00598.

%%%%%%%%%%%%%%%%%%%%%%%%%%%%%%%%%%%%%%%%%%%%%%%%%%%%%%%%%%%%%%%%%%%%%%
%
%
\section*{Data availability}
The data that support the findings of this study are available from the corresponding author (KC) upon reasonable request.

\def\url#1{}


\begin{thebibliography}{10}

\bibitem{horsthemke1984}
W. Horsthemke and R. Lefever, {\em Noise-inducted transitions. Theory and
  applications in physics, chemistry, and biology} (Springer Verlag, Berlin,
  1984).

\bibitem{devoret1984}
M.~H. Devoret, J.~M. Martinis, D. Esteve, and J. Clarke, Phys. Rev. Lett. {\bf
  53},  1260  (1984).

\bibitem{doering1992}
C.~R. Doering and J.~C. Gadoua, Phys. Rev. Lett. {\bf 69},  2318  (1992).

\bibitem{benzi1981}
R. Benzi, A. Sutera, and A. Vulpiani, J. Phys. A: Math. Gen. {\bf 14},  L453
  (1981).

\bibitem{mantegna1995}
R. Mantegna and B. Spagnolo, Nuovo Cimento D {\bf 17},  873  (1995).

\bibitem{gammaitoni1998}
L. Gammaitoni, P. H\"anggi, P. Jung, and F. Marchesoni, Rev. Mod. Phys. {\bf
  70},  223  (1998).

\bibitem{magnasco1993}
M.~O. Magnasco, Phys. Rev. Lett. {\bf 71},  1477  (1993).

\bibitem{reimann2002}
P. Reimann, Phys. Rep. {\bf 361},  57  (2002).

\bibitem{agudov2001}
N.~V. Agudov and B. Spagnolo, Phys. Rev. E {\bf 64},  035102  (2001).

\bibitem{dubkov2004}
A.~A. Dubkov, N.~V. Agudov, and B. Spagnolo, Phys. Rev. E {\bf 69},  061103
  (2004).

\bibitem{valenti2015}
D. Valenti, L. Magazz\`u, P. Caldara, and B. Spagnolo, Phys. Rev. B {\bf 91},
  235412  (2015).

\bibitem{gammaitoni1995}
L. Gammaitoni, F. Marchesoni, and S. Santucci, Phys. Rev. Lett. {\bf 74},  1052
   (1995).

\bibitem{valenti2014}
D. Valenti, C. Guarcello, and B. Spagnolo, Phys. Rev. B {\bf 89},  214510
  (2014).

\bibitem{spagnolo2015}
B. Spagnolo, D. Valenti, C. Guarcello, A. Carollo, D. {Persano Adorno}, S.
  Spezia, N. Pizzolato, and B. {Di Paola}, Chaos Solitons Fractals {\bf 81},
  412   (2015), the emergence of self-organization in complex systems.

\bibitem{spagnolo2017nonlinear}
B. Spagnolo, C. Guarcello, L. Magazz{\`u}, A. Carollo, D. Persano~Adorno, and
  D. Valenti, Entropy {\bf 19},  20  (2017).

\bibitem{dubkov2009}
A.~A. Dubkov, A.~L. Cognata, and B. Spagnolo, J. Stat. Mech. {\bf 2009},
  P01002  (2009).

\bibitem{pankratov2000}
A.~L. Pankratov and M. Salerno, Phys. Lett. A {\bf 273},  162  (2000).

\bibitem{dybiec2008d}
B. Dybiec, E. Gudowska-Nowak, and I.~M. Sokolov, Phys. Rev. E {\bf 78},  011117
   (2008).

\bibitem{lisowski2015}
B. Lisowski, D. Valenti, B. Spagnolo, M. Bier, and E. Gudowska-Nowak, Phys.
  Rev. E {\bf 91},  042713  (2015).

\bibitem{solomon1993}
T.~H. Solomon, E.~R. Weeks, and H.~L. Swinney, Phys. Rev. Lett. {\bf 71},  3975
   (1993).

\bibitem{barthelemy2008}
P. Barthelemy, J. Bertolotti, and D. Wiersma, Nature (London) {\bf 453},  495
  (2008).

\bibitem{mercadier2009levyflights}
M. Mercadier, W. Guerin, M. M.~Chevrollier, and R. Kaiser, Nat. Phys. {\bf 5},
  602  (2009).

\bibitem{cabrera2004}
J.~L. Cabrera and J.~G. Milton, Chaos {\bf 14},  691  (2004).

\bibitem{bouchaud1990}
J.~P. Bouchaud and A. Georges, Phys. Rep. {\bf 195},  127  (1990).

\bibitem{bouchaud1991}
J.~P. Bouchaud, A. Ott, D. Langevin, and W. Urbach, J. Phys. II France {\bf 1},
   1465  (1991).

\bibitem{laherrere1998}
J. Laherr{\`e}re and D. Sornette, Eur. Phys. J. B {\bf 2},  525  (1998).

\bibitem{mantegna2000}
R.~N. Mantegna and H.~E. Stanley, {\em An introduction to econophysics.
  Correlations and complexity in finance} (Cambridge University Press,
  Cambridge, 2000).

\bibitem{lera2018gross}
S.~C. Lera and D. Sornette, Phys. Rev. E {\bf 97},  012150  (2018).

\bibitem{brockmann2006}
D. Brockmann, L. Hufnagel, and T. Geisel, Nature (London) {\bf 439},  462
  (2006).

\bibitem{sims2008}
D.~W. Sims, E.~J. Southall, N.~E. Humphries, G.~C. Hays, C.~J.~A. Bradshaw,
  J.~W. Pitchford, A. James, M.~Z. Ahmed, A.~S. Brierley, M.~A. Hindell, D.
  Morritt, M.~K. Musyl, D. Righton, E.~L.~C. Shepard, V.~J. Wearmouth, R.~P.
  Wilson, M.~J. Witt, and J.~D. Metcalfe, Nature (London) {\bf 451},  1098
  (2008).

\bibitem{barkai2014}
E. Barkai, E. Aghion, and D.~A. Kessler, Phys. Rev. X {\bf 4},  021036  (2014).

\bibitem{amor2016}
T.~A. Amor, S.~D.~S. Reis, D. Campos, H.~J. Herrmann, and J.~S. Andrade, Sci.
  Rep. {\bf 6},  20815  (2016).

\bibitem{shlesinger1986}
M.~F. Shlesinger and J. Klafter,  in {\em On Growth and Form: Fractal and
  Non-fractal Patterns in Physics}, edited by H.~E. Stanley and N. Ostrowsky
  (Springer Verlag, Berlin, 1986), p.\ 279.

\bibitem{reynolds2009}
A.~M. Reynolds and C.~J. Rhodes, Ecology {\bf 90},  877  (2009).

\bibitem{getoor1961}
R.~K. Getoor, Trans. Am. Math. Soc. {\bf 101},  75  (1961).

\bibitem{blumenthal1961}
R.~M. Blumenthal, R.~K. Getoor, and D.~B. Ray, Trans. Am. Math. Soc. {\bf 99},
  540  (1961).

\bibitem{metzler2000}
R. Metzler and J. Klafter, Phys. Rep. {\bf 339},  1  (2000).

\bibitem{barkai2001}
E. Barkai, Phys. Rev. E {\bf 63},  046118  (2001).

\bibitem{chechkin2006}
A.~V. Chechkin, V.~Y. Gonchar, J. Klafter, and R. Metzler,  in {\em Fractals,
  Diffusion, and Relaxation in Disordered Complex Systems: Advances in Chemical
  Physics, Part B}, edited by W. T.~Coffey and Y. P.~Kalmykov (John Wiley \&
  Sons, New York, 2006), Vol.~133, pp.\ 439--496.

\bibitem{jespersen1999}
S. Jespersen, R. Metzler, and H.~C. Fogedby, Phys. Rev. E {\bf 59},  2736
  (1999).

\bibitem{klages2008}
R. Klages, G. Radons, and I.~M. Sokolov, {\em Anomalous transport: Foundations
  and applications} (Wiley-VCH, Weinheim, 2008).

\bibitem{dubkov2008}
A.~A. Dubkov, B. Spagnolo, and V.~V. Uchaikin, Int. J. Bifurcation Chaos. Appl.
  Sci. Eng. {\bf 18},  2649  (2008).

\bibitem{Dubkov2008b}
A.~A. Dubkov and B. Spagnolo, Eur. Phys. J. B {\bf 65},  361  (2008).

\bibitem{Guarcello2019}
C. Guarcello, D. Valenti, B. Spagnolo, V. Pierro, and G. Filatrella, Phys. Rev.
  Applied {\bf 11},  044078  (2019).

\bibitem{brockmann2002}
D. Brockmann and I.~M. Sokolov, Chem. Phys. {\bf 284},  409  (2002).

\bibitem{wilk2000}
G. Wilk and Z. W{\l}odarczyk, Phys. Rev. Lett. {\bf 84},  2770  (2000).

\bibitem{tsallis1995}
C. Tsallis, S.~V.~F. Levy, A.~M.~C. Souza, and R. Maynard, Phys. Rev. Lett {\bf
  75},  3589  (1995).

\bibitem{beck2001}
C. Beck, Phys. Rev. Lett. {\bf 87},  180601  (2001).

\bibitem{beck2003}
C. Beck and E.~G.~D. Cohen, Physica A {\bf 322},  267  (2003).

\bibitem{sparre1953}
E. Sparre~Andersen, Math. Scand. {\bf 1},  263  (1953).

\bibitem{sparre1954}
E. Sparre~Andersen, Math. Scand. {\bf 2},  195  (1954).

\bibitem{molini2011first}
A. Molini, P. Talkner, G.~G. Katul, and A. Porporato, Physica A {\bf 390},
  1841  (2011).

\bibitem{dybiec2008e}
B. Dybiec, Phys. Rev. E {\bf 78},  061120  (2008).

\bibitem{dybiec2009e}
B. Dybiec, Phys. Rev. E {\bf 80},  041111  (2009).

\bibitem{cox1965}
D.~R. Cox and H.~D. Miller, {\em The theory of stochastic processes} (Chapman
  and Hall, London, 1965).

\bibitem{gardiner2009}
C.~W. Gardiner, {\em Handbook of stochastic methods for physics, chemistry and
  natural sciences} (Springer Verlag, Berlin, 2009).

\bibitem{dybiec2010c}
B. Dybiec, Acta Phys. Pol. B {\bf 41},  1127  (2010).

\bibitem{dybiec2012fractional}
B. Dybiec and E. Gudowska-Nowak,  in {\em Fractional dynamics: recent
  advances}, edited by J. Klafter, S.~T. Lim., and R. Metzler (World Scientific
  Publishing, Singapore, 2012), pp.\ 33--50.

\bibitem{katzav2008spectrumfractional}
E. Katzav and M. Adda-Bedia, EPL (Europhys. Lett.) {\bf 83},  30006  (2008).

\bibitem{zozor2011spectral}
S. Zozor and C. Vignat, Phys. Rev. E {\bf 84},  031115  (2011).

\bibitem{kwasnicki2012eigenvalues}
M. Kwa{\'s}nicki, J. Funct. Anal. {\bf 262},  2379  (2012).

\bibitem{kirichenko2016}
E.~V. Kirichenko, P. Garbaczewski, V. Stephanovich, and M.
  \ifmmode~\dot{Z}\else \.{Z}\fi{}aba, Phys. Rev. E {\bf 93},  052110  (2016).

\bibitem{ditlevsen1999}
P.~D. Ditlevsen, Phys. Rev. E {\bf 60},  172  (1999).

\bibitem{bier2018}
M. Bier, Phys. Rev. E {\bf 97},  022113  (2018).

\bibitem{imkeller2006}
P. Imkeller and I. Pavlyukevich, Stoch. Proc. Appl. {\bf 116},  611  (2006).

\bibitem{imkeller2006b}
P. Imkeller and I. Pavlyukevich, J. Phys. A: Math. Gen. {\bf 39},  L237
  (2006).

\bibitem{chechkin2003b}
A.~V. Chechkin, R. Metzler, V.~Y. Gonchar, J. Klafter, and L.~V. Tanatarov, J.
  Phys. A: Math. Gen. {\bf 36},  L537  (2003).

\bibitem{koren2007}
T. Koren, M.~A. Lomholt, A.~V. Chechkin, J. Klafter, and R. Metzler, Phys. Rev.
  Lett. {\bf 99},  160602  (2007).

\bibitem{koren2007b}
T. Koren, A.~V. Chechkin, and J. Klafter, Physica A {\bf 379},  10  (2007).

\bibitem{palyulin2019first}
V.~V. Palyulin, G. Blackburn, M.~A. Lomholt, N.~W. Watkins, R. Metzler, R.
  Klages, and A.~V. Chechkin, New J. Phys. {\bf 21},  103028  (2019).

\bibitem{janicki1994b}
A. Janicki and A. Weron, Stat. Sci. {\bf 9},  109  (1994).

\bibitem{samorodnitsky1994}
G. Samorodnitsky and M.~S. Taqqu, {\em Stable non-{Gaussian} random processes:
  Stochastic models with infinite variance} (Chapman and Hall, New York, 1994).

\bibitem{dybiec2017levy}
B. Dybiec, E. Gudowska-Nowak, E. Barkai, and A.~A. Dubkov, Phys. Rev. E {\bf
  95},  052102  (2017).

\bibitem{janicki1994}
A. Janicki and A. Weron, {\em Simulation and chaotic behavior of
  $\alpha$-stable stochastic processes} (Marcel Dekker, New York, 1994).

\bibitem{janicki1996}
A. Janicki, {\em Numerical and statistical approximation of stochastic
  differential equations with {non-Gaussian} measures} (Hugo Steinhaus Centre
  for Stochastic Methods, Wroc{\l}aw, 1996).

\bibitem{dybiec2018conservative}
B. Dybiec, K. Capa{\l}a, A.~V. Chechkin, and R. Metzler, J. Phys. A: Mat.
  Theor. {\bf 52},  015001  (2019).

\bibitem{sokolov2004distributed}
I.~M. Sokolov, A.~V. Chechkin, and J. Klafter, Acta Phys. Pol. B {\bf 34},
  1323  (2004).

\bibitem{meerschaert2011distributed}
M.~M. Meerschaert, E. Nane, and P. Vellaisamy, J. Math. Anal. Appl. {\bf 379},
  216  (2011).

\end{thebibliography}
\end{document}